\documentclass[twocolumn,showpacs,preprintnumbers,superscriptaddress,
amsmath,amssymb,UTF8]{revtex4-2}

\usepackage{epsfig}
\usepackage{graphicx}
\usepackage{dcolumn}
\usepackage{bm}
\usepackage{times}
\usepackage{xcolor}
\usepackage{amsthm}
\usepackage{color}
\usepackage{epstopdf}
\usepackage{mathtools}

\begin{document}
\title{Competing spreading dynamics in simplicial complex}

\author{WenYao Li}
\affiliation{College of Computer Science, Sichuan University, Chengdu 610065,
China}

\author{Xiaoyu Xue}
\affiliation{School of Cyber Science and Engineering, Sichuan University, Chengdu 610065,
China}

\author{Liming Pan}
\affiliation{School of Computer and Electronic Information, Nanjing Normal
University - Nanjing 210023, China}

\author{Tao Lin}\email{lintao@scu.edu.cn}
\affiliation{College of Computer Science, Sichuan University, Chengdu 610065,
China}

\author{Wei Wang}\email{wwzqbx@hotmail.com}
\affiliation{Cybersecurity Research Institute, Sichuan University, Chengdu
610065, China}
\date{\today}
\begin{abstract}
Interactions in biology and social systems are not restricted to pairwise but can take arbitrary sizes. Extensive studies have revealed that the arbitrary-sized interactions significantly affect the spreading dynamics on networked systems. Competing spreading dynamics, i.e., several epidemics spread simultaneously and compete with each other, have been widely observed in the real world, yet the way arbitrary-sized interactions affect competing spreading dynamics still lacks systematic study.  This study presents a model of two competing simplicial susceptible-infected-susceptible epidemics on a higher-order system represented by simplicial complex and analyzes the model's critical phenomena. In the proposed model, a susceptible node can only be infected by one of the two epidemics,  and the transmission of infection to neighbors can occur through pairwise (i.e.,  an edge) and high-order (e.g., 2-simplex) interactions simultaneously.  Through a mean-field (MF) theory analysis and numerical simulations, we show that the model displays rich dynamical behavior depending on the 2-simplex infection strength.  When the 2-simplex infection strength is weak, the model's phase diagram is consistent with the simple graph, consisting of three regions: the absolute dominant regions for each epidemic and the epidemic-free region. With the increase of the 2-simplex infection strength, a new phase region called the alternative dominant region emerges. In this region, the survival of one epidemic depends on the initial conditions. Our theoretical analysis can reasonably predict the time evolution and steady-state outbreak size in each region. In addition, we further explore the model's phase diagram both when the 2-simplex infection strength is symmetrical and asymmetrical. The results show that the 2-simplex infection strength has a significant impact on the system phase diagram.
\end{abstract}
\maketitle

\section{Introduction}
Competitive epidemic spreading dynamics has received widespread attention in network science, physics, and mathematics~\cite{Wang2019,Pan2020}, as it describes various spreading processes in real-world systems. For competing spreading dynamics such as two competing epidemics, a host can be infected by only one of the two, since the epidemic kills the host before infection by the second can occur, or there may be cross-immunity between the two epidemics~\cite{Newman2005}. Besides epidemics, competing spreading can also model products flowing in the market. After choosing a product, consumers will lose interest in other similar products for a period of time~\cite{Dubey2006}. Similarly, for computer viruses spreading on the Internet, after people realize that a computer virus infects their computer, they are likely to install anti-virus software to kill the virus, making it less likely to be infected by other viruses~\cite{Newman2002,marceau2011modeling}. Studies in the literature have been focusing on revealing the essential dynamical properties of competing dynamics, such as whether two epidemics can coexist in the steady-state, and if not, which epidemic will eventually survive; the phase transition between different steady-states, as well as the type of phase transitions~\cite{Wang2019}. Studying competing spreading dynamics can provide insight into the intervention of real-world spreading processes.  For epidemic prevention, from understandings the epidemic transmission mechanism, the government could adjust and optimize the epidemic prevention strategies~\cite{Brockmann2013,Gallotti2020}. Besides, companies can do competitive product analyses to help designing product operation strategy~\cite{Mahajan1991,Prakash2012,Valera2015}. In addition, for finance, the studies can help avoid financial risks and discovering financial opportunities~\cite{Claessens2001,Helbing2013}.

Previous studies have extensively analyzed the way spreading mechanisms and underlying network topology affect the competing spreading dynamics. Newman~\cite{Newman2005} studied two susceptible-infected-removed (SIR) competing epidemic spreading on networks using bound percolation theory and showed that it is possible to observe the coexistence of two SIR epidemics. The results have raised lots of further discussions~\cite{Newman2013,Karrer2011,Poletto2015}. Prakash et al.~\cite{Prakash2012} conducted a theoretical analysis for the full mutual immunity model on arbitrary topology and proved that the `winner takes all,' i.e., the more potent product will hold the dominance, and the weaker product will become extinct. They further studied the problem of coexistence in the SIS model of partial competition~\cite{Beutel2012}. Wu et al.~\cite{Wu2013} studied two SIS epidemics with different reproductive numbers that spread on scale-free networks and found the coexistence of the two epidemics.  Li et al.~\cite{Li2018} studied two mutually reinforcing epidemics spread under the limit of resources and find the critical value of resources that inhibit the spread of these two epidemics. The interaction between epidemics can not only be competitive but also be promotional~\cite{Mills2013,Rehman2021,BasslerBonnie2014,CaiWeiran2015,Liu2018}, or asymmetrical~\cite{Wang2021}.
When considering multi-layer networks, the competitive spreading problem becomes more complex~\cite{Cozzo2018}. Funk and Jansen~\cite{Funk2010} studied bond percolation of two different processes on overlay networks of arbitrary joint degree distribution. Faryad et al.~\cite{DarabiSahneh2014} discussed the spread of two competing viruses in host populations with different contact networks from a comprehensive topology perspective.

Most of the studies are based on simple graphs which regard individuals as nodes and the relationship between individuals as connected edges. The fundamental limitation of simple graphs is that it only captures pairwise interactions, while many systems display group interactions~\cite{CentolaD2010,Ugander2012,Weng2012,
Karsai2014,Monsted2017,Guilbeault2017,Battiston2020}. For instance, scientific research is often carried out by a group of people, and the contagion of rumors or spreading of ideas can take place in the form of groups. The importance of higher-order interactions has been realized for a long time~\cite{Atkin1972,Welsh1974,Guilbeault2017}. Facing the challenges of modeling higher-order interactions, scientists tried to use pairwise interactions to approximate group interactions. For example, there has been studies using bipartite graph~\cite{newman2001random} or clique expansion~\cite{Dunbar1995,Derenyi2005,Palla2005,Kahle2009,AnnSizemore2018,Wang2018} to model higher-order interaction, however the results were not satisfactory.

Researchers have been committed to designing a proper mathematical framework for describing group interactions in a natural way~\cite{Carlsson2009,Patania2017}. Simplicial complex~\cite{Kee2013,Iacopini2019} describes the higher-order interactions by interaction sets rather than pairwise edges. If a simplex $\sigma$ included in simplicial complex $\kappa$, then all the sub-simplices $v \subset \sigma$ of simplex $\sigma$ are also contained in $\kappa$. Iacopini et al.~\cite{Iacopini2019} proposed a higher-order model of social contagion on simplicial complex and found a discontinuous phase transition and bistable region in the phase diagram. Through a mean-field analysis, they found that the 2-simplex infection strength decides the discontinuous transition, and the steady-state in the bistable region relies on the fraction of initial seeds. The microscopic Markov chain approach~\cite{Gomez2010} and the epidemic link equations~\cite{Matamalas2018} have been adopted in improving the accuracy of mean-field approaches. Matamalas et al. obtained a more accurate prediction of the spreading dynamics on simplicial complex~\cite{Matamalas2020}. Compared to simplicial complex,  hypergraphs~\cite{Higuchi1999,Ghoshal2009,Kumar2018,Jhun2019,Chodrow2019} do not require the appearance of all subsets in each interaction set, thus is more flexible in describing higher-order interactions. Hypergraph models, such as the uniform hypergraph, have been proposed to describe the higher-order interactions and to investigate the dynamics of group epidemic spreading~\cite{Bodo2016,Landry2020}.

From previous studies, higher-order interactions have an essential effect on the spread of a single epidemic. To the best of our knowledge, there still lacks theoretical studies of its influence on the competitive spreading dynamics when two epidemics spreading on the network simultaneously. This study proposes an absolute competing model for two SIS-type epidemics that are homogeneously mixed on the simplicial complex with a highest interaction dimension $D=2$. In Sec.~\ref{sec:model}, we introduce the competing spreading dynamics model on the simplicial complex. Then, we derive the MF theory rate equations in Sec.~\ref {sec:Theoretical results}, which includes both interactions of the first and second order. We obtain seven fixed points of the rate equation and analyze their stability of the system. In Sec.~\ref{sec:Results analyses} we study the conditions for the fixed points to be stable and obtain the phase diagram of the proposed model. When the 2-simplex infection strength is weak, the phase space has three regions, similar to simple graphs. The regions include the absolute dominant regions for each epidemic and the epidemic-free region. With the increase of the 2-simplex infection strength, alternative dominant regions emerge, in which the fraction of the initial seeds decide the survival of the epidemic. In this case, the phase diagram has into nine regions with the hysteresis loop appears. Combining both the theory and extensive numerical simulations, we illustrate the evolution process of epidemics in each region by evolution diagram and predict the outbreak size of the epidemics. The results show that the theory and simulation agree well. We further discuss the influence of the 2-simplex infection strength on the phase diagram. When the 2-simplex infection strength of epidemics is symmetrical and asymmetrical, the existence and size of alternate dominant regions are related to the 2-simplex infection strength. Finally, we present conclusions and discussions in Sec.~\ref{sec:Conclusions and Discussions}.

\section{Model descriptions} \label{sec:model}
In this section, we propose a competing spreading dynamics on simplicial complex $H=(V,\kappa)$, where $V$ denotes the node set, $\kappa$ stands for the $k$-simplex set. A $k$-simplex $\sigma\in\{\kappa\}$ is defined as the interactions of a set of $k + 1$ vertices $\sigma = [ v_0 , . . . , v_k ]$. Therefore, 0-simplex is one single node, and 1-simplex represents two nodes, and 2-simplex is the collection of three nodes, and so on. There is an extra requirement for simplicial complex that if a simplex $\sigma \in \kappa$, then all the sub-simplices $v \subset \sigma$ of simplex $\sigma$ are also contained in $\kappa$. For example, a 2-simplex is consists of three nodes $\sigma=[v_0,v_1,v_2] \in \kappa$, whose subsimplices $[v_0]$, $[v_1]$, $[v_2]$, $[v_0,v_1]$, $[v_0,v_2]$ and $[v_1,v_2]$ are also belong to $\kappa$.

We use random simplicial complex (RSC) model~\cite{Iacopini2019} to generate the artificial simplicial complex. The RSC allows us to generate the simplicial complex with specified average degree to each dimension. To generate the $D$ dimension simplicial complex, we need $D+1$ parameters that is $N$ vertices and $D$ probabilities $\{p_1,...,p_D\}$ whose elements control the creation of simplices in each dimension. In this paper, we set $D = 2$. The RSC model can be generated as follows. Given a set $V$ with $N$ vertices, we first connect the pair of nodes first with probability $p_1=(\langle k \rangle-2\langle k^{\prime} \rangle)/(N-1-2\langle k^{\prime} \rangle)$ (i.e., the 1-simplex connecting probability), where $\langle k\rangle$ is the average degree of the 1-simplex, and $\langle k^{\prime}\rangle$ means the average degree of the 2-simplex. Next, randomly select three vertices with probability $p_2= 2\langle k^{\prime} \rangle /[(N-1)(N-2)]$ to create 2-simplex. The average degree of the simplicial complex is $\langle k\rangle=(N-1)p_1+2\langle k^{\prime}\rangle(1-p_1)$.

\begin{figure}[]
\begin{center}
\epsfig{file=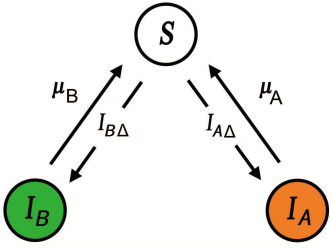,width=0.5\linewidth}
\caption{State transformation on competitive simplicial complex. In the simplex where the infection of epidemic $X \in \{A,B\}$ is infected, susceptible individuals become $X$-infected with rate $I_{X\Delta}$, and $X$-infected individuals recover to susceptible with rate $\mu_X$.}
\label{fig1}
\end{center}
\end{figure}

Consider the absolute competition between epidemic A and epidemic B in the simplicial complex. If the host has disease A, it will not be infected by epidemic B or vice versa. For each spreading dynamics, we assume that it follows a simplicial susceptible-infected-susceptible (SIS) model, which is proposed in Ref.~\cite{Iacopini2019}.
A node can transform among three states: susceptible state $S$, A-infected state $I_A$, and B-infected state $I_B$. $S$ state node can transform to $I_A$ state with probability $I_{A\Delta}$, and becomes $I_B$ state with probability $I_{B\Delta}$. An $I_A$ ($I_B$) state node recovers to $S$ state by itself with probability $\mu_A$ ($\mu_B$). The above state transition is illustrated in Fig.~\ref{fig1}.

Epidemic $X \in \{A,B\}$ spreading on simplicial complex is governed by $2$ control parameters $I_{X\Delta}\in\{\beta_{X1},\beta_{X2}\}$, $\beta_{X1}$ describes the 1-simplex infectivity rate (pairwise interaction), and $\beta_{X2}$ describes the 2-simplex infectivity rate (2-order interaction). If a susceptible node $i$ and $X$-infected nodes $j$ are connected by a 1-simplex, node $i$ obtains the infection through this 1-simplex $j\rightarrow i$ with rate $\beta_{X1}$. When a susceptible node $i$ and two nodes $j$ and $\ell$ are connected by a 2-simplex, there are two situations. (i) If one of nodes $j$ and $\ell$ is in the $X$-infected state, node $i$ can only get the infection through 1-simplex with rate $\beta_{X1}$. (ii) If nodes $j$ and $\ell$ are in the $X$-infected state, node $i$ will get the infection from $j$ and $\ell$ with rate $1-(1-\beta_{X1})^2$ through the two pieces of 1-simplex connected between $j\rightarrow i$ and $\ell\rightarrow i$. In addition, node $i$ also get an addition infection rate through the 2-simplex with rate $\beta_{X2}$. Therefore, the infection probability of node $i$ in situation (ii) is $1-(1-\beta_{X1})^2(1-\beta_{X2})$.

\section{Theoretical results} \label{sec:Theoretical results}
In this section, we first develop a MF theory and then analyze the stability of the system.

\par %
\subsection{Rate equations}
Let us denote $s(t)$, $\rho_A(t)$ and $\rho_B(t)$ as the fraction of susceptible state, A-infected state and B-infected state, they satisfy the conservation function that $s(t)+\rho_A(t)+\rho_B(t)=1 $. The dynamics of the model can be described as MF equations, in which we assume that there are no statistical differences among distinct nodes. That is to say, for any node $i$ and $j$, $i\neq j$, have the same probability in the same states. The rate equations of infected node fraction of two epidemics are as\begin{align}
&\begin{split}
\label{eq:eq1_1}
d_t\rho_A(t)=&-\mu_A\rho_A(t)\\
& +\sum_{w=1}^{D} \beta_{Aw}\langle k_w\rangle  \rho_A^w(t) [1-\rho_A(t)-\rho_B(t)],
\end{split}
\\
&\begin{split}
\label{eq:eq1_2}
d_t\rho_B(t)=&-\mu_B\rho_B(t)\\
&+\sum_{w=1}^{D} \beta_{Bw}\langle k_w\rangle  \rho_B^w(t) [1-\rho_A(t)-\rho_B(t)],
\end{split}
\end{align}
where $\langle k_w\rangle $ represents the average degree of the $w$-simplex. The first term on the right side of Eq.~(\ref{eq:eq1_1}) is the reduction of evolution rate for epidemic A due to the recovery. The second term as the gain term of evolution rate of epidemic A, represents the fraction of nodes newly infected by epidemic A in each $w$-simplex. Similar to epidemic B, Eq.~(\ref{eq:eq1_2}) describes the evolution rate of epidemic B.

In this paper, we focus on higher-order interactions with $D = 2 $. Eqs. (\ref{eq:eq1_1}) and (\ref{eq:eq1_2}) can be further expressed as
\begin{equation}
\begin{split}
\label{eq:D2A}
d_t\rho_A(t)=&-\mu_A\rho_A(t)
+ \rho_A(t)\beta_{A} \langle k\rangle  [1-\rho_A(t)-\rho_B(t)]
\\&	+\rho_A(t)^2\beta_{A}^{\prime} \langle k^{\prime}\rangle  [1-\rho_A(t)-\rho_B(t)],
\end{split}
\end{equation}
and
\begin{equation}
\begin{split}
\label{eq:D2B}
d_t\rho_B(t)&=-\mu_B\rho_B(t)
+\rho_B(t) \beta_{B} \langle k\rangle  [1-\rho_A(t)-\rho_B(t)]\\
&+\rho_B(t)^2\beta_{B}^{\prime} \langle k^{\prime}\rangle  [1-\rho_A(t)-\rho_B(t)],
\end{split}
\end{equation}
respectively. For simplicity, we use $\beta_A$, $\beta_B$ and $\langle k\rangle $ represent the $1$-simplex (i.e., pairwise interaction), and use $\beta_{A}^{\prime}$, $\beta_{B}^{\prime}$ and $\langle k^{\prime}\rangle $ represent the $2$-simplex.

For the epidemic spreading dynamics, an important parameter is the basic reproductive number $R_0$ \cite{Dietz1993,Guerra2017,Pastor-satorras2015}, representing the average number of new infections triggered by an infected node. When $R_0>1$, a global epidemic may break out; otherwise, no epidemic exists in the system. For the case of homogeneous population or networks, we know $R_0=\beta\langle k\rangle /\mu$, where $\beta$ is the infection rate, $\mu$ denotes the recovery rate, and $\langle k\rangle $ is the average degree of the homogeneous population. With the denotation of $R_0$, we integrate three parameters into $R_0$. Denoting $\lambda_A=\beta_A \langle k \rangle /\mu_A$ and $\lambda_A^{\prime}=\beta_{A}^{\prime} \langle k^{\prime} \rangle /\mu_A$ represent the basic reproductive number when only 1-simplex and 2-simplex include for epidemic A, respectively.
Similarly, we denote $\lambda_B=\beta_B \langle k \rangle /\mu_B$ and $\lambda_B^{\prime}=\beta_{B}^{\prime} \langle k^{\prime} \rangle /\mu_B$ respectively stand for the basic reproductive number
(i.e., the infection strength) when only 1-simplex and 2-simplex include for epidemic B. We know the larger value of $\lambda_A^\prime $, the stronger of the 2-simplex infectivity. Other parameters have the similar meanings. Eqs.~(\ref{eq:D2A}) and (\ref{eq:D2B}) can be further simplified as
\begin{equation}
\begin{split}
\label{eq:rate_eq_A}
d_t\rho_A(t) =&-\mu_A\rho_A(\lambda_{A}^{\prime}\rho_A^2+			(\lambda_{A}-\lambda_{A}^{\prime}+\lambda_{A}^{\prime}\rho_B)\rho_A
\\&+(1-\lambda_{A}+\lambda_{A}\rho_B)),
\end{split}
\end{equation}
and		
\begin{equation}
\begin{split}
\label{eq:rate_eq_B}
d_t\rho_B(t)=& -\mu_B\rho_B(\lambda_{B}^{\prime}\rho_B^2+
(\lambda_{B}-\lambda_{B}^{\prime}+\lambda_{B}^{\prime}\rho_A)\rho_B
\\&+(1-\lambda_{B}+\lambda_{B}\rho_A)),
\end{split}
\end{equation}
respectively.

\subsection{Stability analyses}
\begin{figure}[ht]
\begin{center}
\epsfig{file=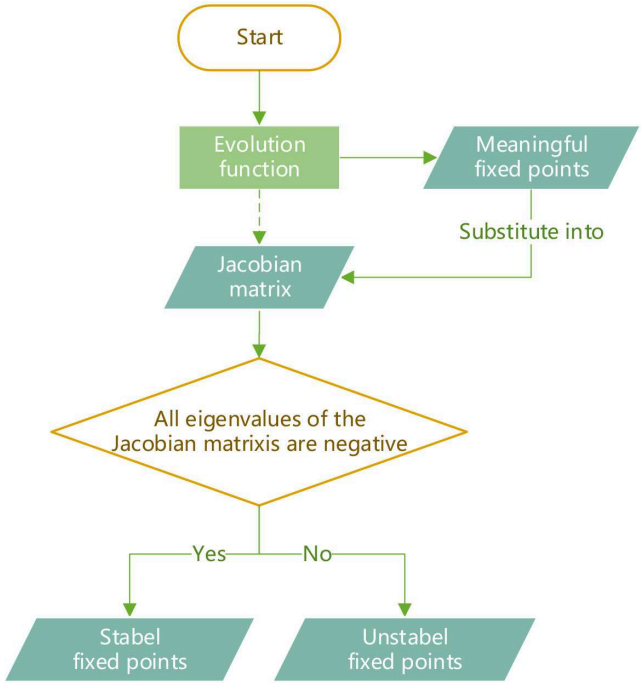,width=\linewidth}
\caption{Illustration of the procedures for stability of the fixed points.}
\label{fig2}
\end{center}
\end{figure}

When $t\rightarrow \infty$, i.e., $d_t\rho_A(t)=0$ and $d_t\rho_B(t)=0$, the system reaches a dynamical steady state. By setting the left hand of Eqs.~(\ref{eq:rate_eq_A}) and (\ref{eq:rate_eq_B}) equal to zero, we can solving the fixed points of the competing spreading dynamics. With the knowledge of nonlinear dynamics, the procedures for stability of the fixed points are illustrated in Fig.~\ref{fig2}.
We find 7 possible fixed points $(\rho_A^*,\rho_B^*)$ for the given values of dynamical parameters. The seven fixed points are as follows.
\begin{itemize}
\item Fixed point 1: $(\rho_A^*,\rho_B^*)=(0,0)$, which means that two epidemic die-out.
\item Fixed point 2: $(\rho_A^*,\rho_B^*)=(\rho_{A-}^*,0)$, which indicates that only epidemic A survives, where
\begin{equation}\rho_{A-}^*=\frac{{\lambda_A^{\prime}-\lambda_A} - \sqrt{\lambda_A ^ 2 + 2 \lambda_A {\lambda_{A}^{\prime}} + {\lambda_{A}^{\prime}} ^ 2 - 4 {\lambda_{A}^{\prime}}}}{2 {\lambda_{A}^{\prime}}}.
\end{equation}
\item Fixed point 3: $(\rho_A^*,\rho_B^*)=(\rho_{A+}^*,0)$, represents that only epidemic A survives, and the steady fraction of epidemic A is
\begin{equation}\rho_{A+}^*=\frac{{\lambda_A^{\prime}-\lambda_A} + \sqrt{\lambda_A ^ 2 + 2 \lambda_A {\lambda_{A}^{\prime}} + {\lambda_{A}^{\prime}} ^ 2 - 4 {\lambda_{A}^{\prime}}}}{2 {\lambda_{A}^{\prime}}}.
\end{equation}
\item Fixed point 4: $(\rho_A^*,\rho_B^*)=(0,\rho_{B-}^*)$, only when epidemic B survives, where
\begin{equation} \rho_{B-}^*=\frac{ {\lambda_{B}^{\prime}} -\lambda_B - \sqrt{\lambda_B ^ 2 + 2 \lambda_B {\lambda_{B}^{\prime}} + {\lambda_{B}^{\prime}} ^ 2 - 4 {\lambda_{B}^{\prime}}} }{2 {\lambda_{B}^{\prime}}}.
\end{equation}

\item Fixed point 5: $(\rho_A^*,\rho_B^*)=(0,\rho_{B+}^*)$, which represents that only epidemic B survives, where the steady fraction of epidemic B is
\begin{equation} \rho_{B+}^*=\frac{ {\lambda_{B}^{\prime}} -\lambda_B + \sqrt{\lambda_B ^ 2 + 2 \lambda_B {\lambda_{B}^{\prime}} + {\lambda_{B}^{\prime}} ^ 2 - 4 {\lambda_{B}^{\prime}}} }{2 {\lambda_{B}^{\prime}}}.\end{equation}

\item Fixed point 6: $(\rho_A^*,\rho_B^*)=(\rho_{A_6}^*, \rho_{B_6}^*)$, which indicates two epidemic coexistence, where
\begin{equation}
\label{fixed_point_6}
\begin{split}
\left\{
\begin{array}{lr}
\rho_{A_6}^*=\frac{\lambda_B \lambda_A^{\prime} - \lambda_A \lambda_B^{\prime} - 2 \lambda_A \lambda_A^{\prime} + \lambda_A^{\prime} \lambda_B^{\prime} + \sqrt{\phi_1}}{2\lambda_A^\prime(\lambda_A^\prime + \lambda_B^{\prime})}
\\
\rho_{B_6}^*=\frac{ \lambda_A \lambda_B^{\prime} - \lambda_B \lambda_A^{\prime} - 2 \lambda_B \lambda_B^{\prime} + \lambda_A^{\prime} \lambda_B^{\prime} + \sqrt{\phi_1}}{2\lambda_B^\prime(\lambda_A^\prime + \lambda_B^\prime)}
\end{array}
\right.,
\end{split}
\end{equation}
and
\begin{equation}
\begin{split}
		\phi_1=& \lambda_A^2 \lambda_B^{\prime 2} +
		2 \lambda_A \lambda_B \lambda_A^{\prime} \lambda_B^{\prime} +
		2 \lambda_A \lambda_A^{\prime} \lambda_B^{\prime 2} +
		\lambda_B^2 \lambda_A^{\prime 2} + \\&
		2 \lambda_B \lambda_A^{\prime 2} \lambda_B^{\prime} +
		\lambda_A^{\prime 2} \lambda_B^{\prime 2} -
		4 \lambda_A^{\prime 2} \lambda_B^{\prime} -
		4 \lambda_A^{\prime} \lambda_B^{\prime 2}.
\end{split}
\end{equation}

\item Fixed point 7: $(\rho_A^*,\rho_B^*)=(\rho_{A_7}^*, \rho_{B_7}^*)$, which indicates two epidemic coexistence, where
\begin{equation}
\begin{split}
	\left\{
	\begin{array}{lr}
		\rho_{A_7}^*=&\frac{ \lambda_B \lambda_A^{\prime} - \lambda_A \lambda_B^{\prime} - 2 \lambda_A \lambda_A^{\prime} + \lambda_A^{\prime} \lambda_B^{\prime} - \sqrt{\phi_2}}{2\lambda_A^\prime(\lambda_A^\prime + \lambda_B^{\prime})}
				\\
				\rho_{B_7}^*=&\frac{ \lambda_A \lambda_B^{\prime} - \lambda_B \lambda_A^{\prime} - 2 \lambda_B \lambda_B^{\prime} + \lambda_A^{\prime} \lambda_B^{\prime} - \sqrt{\phi_2}}{2\lambda_B^\prime(\lambda_A^\prime + \lambda_B^\prime)}
	\end{array}
	\right.,
	\end{split}
\end{equation}
and,
\begin{equation}
\begin{split}
\phi_2=& \lambda_A^2 \lambda_B^{\prime 2} +
2 \lambda_A \lambda_B \lambda_A^{\prime} \lambda_B^{\prime} +
2 \lambda_A \lambda_A^{\prime} \lambda_B^{\prime 2} +
\lambda_B^2 \lambda_A^{\prime 2}
\\&
+2 \lambda_B \lambda_A^{\prime 2} \lambda_B^{\prime} +
\lambda_A^{\prime 2} \lambda_B^{\prime 2} -
4 \lambda_A^{\prime 2} \lambda_B^{\prime} -
4 \lambda_A^{\prime} \lambda_B^{\prime 2}.
\end{split}
\end{equation}
\end{itemize}

For a given fixed point $(\rho_A^*,\rho_B^*)$, we obtain the corresponding Jacobians matrix from Eqs.~(\ref{eq:rate_eq_A}) and (\ref{eq:rate_eq_B}) as
\begin{align}
J&=\left[
\begin{matrix}
J_{11} & J_{12} \\
J_{21} & J_{22} \\
\end{matrix}
\right],
\end{align}
where
\begin{align}
\begin{split}
J_{11}=& -\mu_A(3\lambda_A^{\prime}{\rho_A^*}^2+2(\lambda_A-\lambda_A^{\prime})\rho_A^*\\
&+(1-\lambda_A)+\lambda_A{\rho_B^*}+2\lambda_A^{\prime}\rho_A^*{\rho_B^*}),
\end{split}		
\\J_{12}=& -\mu_A(\lambda_A\rho_A^*+\lambda_A^{\prime}{\rho_A^*}^2),
\\J_{21}=& -\mu_B(\lambda_B{\rho_B^*}+\lambda_B^{\prime}{\rho_B^*}^2),
\\
\begin{split}
J_{22}=& -\mu_B(3\lambda_B^{\prime}{\rho_B^*}^2+2(\lambda_B-\lambda_B^{\prime}){\rho_B^*}\\
&+(1-\lambda_B)+\lambda_B\rho_A^*+2\lambda_B^{\prime}\rho_A^*{\rho_B^*}).
\end{split}	
\end{align}
The Jacobian matrix $J$ of the dynamic system has two eigenvalues and denote as $\Lambda_1$ and $\Lambda_2$. The system is stable only when the eigenvalues of $J$ are all negative. And the maximum eigenvalue of the Jacobian matrix
\begin{equation}
{\rm max}\{\Lambda_1,\Lambda_2\}=0
\end{equation}
is the critical point of the system.
\subsubsection{Fixed point 1}
Taking the first fixed point $(\rho_A^*,\rho_B^*)\to(0,0)$ into Jacobian matrix, we get the two eigenvalues of $J$ as
\begin{equation}
\Lambda_1=\mu_A(\lambda_A - 1)
\end{equation}
and
\begin{equation}
\Lambda_2=\mu_B(\lambda_B - 1).
\end{equation}
When $\lambda_{A}=\lambda_{B}$ and $\mu_A=\mu_B$, we know $\Lambda_1=\Lambda_2$. The system is stable when $\Lambda_1<0$ and $\Lambda_2<0$, which needs $\lambda_A<1$ and $\lambda_B<1$. When $\Lambda_1>\Lambda_2$, the threshold point is
\begin{equation}
\label{eq:A threshold point of fp1}
\lambda_A^{c}=1.
\end{equation}
Similarly, when $\Lambda_1<\Lambda_2$, the threshold point is
\begin{equation}
\label{eq:B threshold point of fp1}
\lambda_B^{c}=1.
\end{equation}

\subsubsection{Fixed point 2}
When we consider the second fixed point $(\rho_A^*,\rho_B^*)\to(\rho_{A-}^*,0)$, the two eigenvalues of Jacobian matrix are as
\begin{equation}
\Lambda_1=-\mu_B(\lambda_B\rho_{A-}^* - \lambda_B + 1),
\end{equation}
and
\begin{equation}
\Lambda_2= -\mu_A(3\lambda_A^{\prime}{\rho_{A-}^{*2}} + (2\lambda_A - 2\lambda_A^{\prime}){\rho_{A-}^*} - \lambda_A + 1).
\end{equation}
The second fixed point is meaningful when $2\sqrt{\lambda_{A}^\prime}-\lambda_{A}^\prime<\lambda_A<1$. We know $\Lambda_1<0$ when $\lambda_B<1/(1-\rho_{A-}^*)$. However, $\Lambda_2$ is always greater than zero in this valid domain. As a result, the second fixed point is always unstable.

\subsubsection{Fixed point 3}
The eigenvalues of Jacobian matrix $J$ when considering the third fixed point $(\rho_A^*,\rho_B^*)\to(\rho_{A+}^*,0)$ are
\begin{equation}
\Lambda_1=-\mu_B(\lambda_B\rho_{A+}^* - \lambda_B + 1),
\end{equation}
and
\begin{equation}
\Lambda_2= -\mu_A(\phi_3+ 1),
\end{equation}
where
\begin{equation}
\phi_3=3\lambda_A^{\prime}{\rho_{A+}^{*2}} + (2\lambda_A - 2\lambda_A^{\prime}){\rho_{A+}^*} - \lambda_A.
\end{equation}
The third fixed point is meaningful, if $\lambda_A>2\sqrt{\lambda_{A}^\prime}-\lambda_{A}^\prime$.
When $\lambda_B=\phi_3/(\rho_{A+}^*-1)$ and $\mu_A=\mu_B$, we know $\Lambda_1=\Lambda_2$. The system is stable if $\Lambda_1<0$ and $\Lambda_2<0$. When $\Lambda_1>\Lambda_2$, the threshold point is
\begin{equation}
\label{eq:A threshold point of fp3}
\lambda_B^{c_*}=\frac{1}{(1-\rho_{A+}^*)}.
\end{equation}
When $\Lambda_1<\Lambda_2$, the first eigenvalue $\Lambda_1$ will be greater than 0. Thus the fixed point is unstable at this time.

\subsubsection{Fixed point 4}
Taking the fourth fixed point $(\rho_A^*,\rho_B^*)\to(0,\rho_{B-}^*)$ into Jacobian matrix, we get the two eigenvalues of $J$ as
\begin{equation}
\Lambda_1=-\mu_A(\lambda_A\rho_{B-}^* - \lambda_A + 1),
\end{equation}
and
\begin{equation}
\Lambda_2= -\mu_B(3\lambda_B^{\prime}{\rho_{B-}^{*2}} + (2\lambda_B - 2\lambda_B^{\prime}){\rho_{B-}^*} - \lambda_B + 1).
\end{equation}
The fourth fixed point is meaningful for $2\sqrt{\lambda_{B}^\prime}-\lambda_{B}^\prime<\lambda_B<1$. When $\lambda_A<1/(1-\rho_{B-}^*)$, we know $\Lambda_1<0$. However, $\Lambda_2>0$ in this valid domain, which means that the fourth fixed point is always unstable.

\subsubsection{Fixed point 5}
The fifth fixed point is $(\rho_A^*,\rho_B^*)\to(\rho_{B+}^*,0)$, tack it into Jacobian matrix, we get the two eigenvalues of $J$ as
\begin{equation}
\Lambda_1=-\mu_A(\lambda_A\rho_{B+}^* - \lambda_A + 1),
\end{equation}
and
\begin{equation}
\Lambda_2= -\mu_B(\phi_4+ 1),
\end{equation}
where
\begin{equation}
\phi_4=3\lambda_B^{\prime}{\rho_{B+}^{*2}} + (2\lambda_B - 2\lambda_B^{\prime}){\rho_{B+}^*} - \lambda_B.
\end{equation}
The fifth fixed point is meaningful, if $\lambda_B>2\sqrt{\lambda_{B}^\prime}-\lambda_{B}^\prime$.
When $\lambda_A=\phi_4/(\rho_{B+}^*-1)$ and $\mu_A=\mu_B$, we know $\Lambda_1=\Lambda_2$. When $\Lambda_1>\Lambda_2$, the threshold point is
\begin{equation}
\label{eq:B threshold point of fp5}
\lambda_A^{c_*}=\frac{1}{ 1-\rho_{B+}^* }.
\end{equation}
Note that when $\Lambda_1<\Lambda_2$, we know $\Lambda_1>0$, the system is unstable.

\subsubsection{Fixed point 6}
The analytical solutions of the eigenvalues of the fixed point six $(\rho_A^*,\rho_B^*)=(\rho_{A_6}^*, \rho_{B_6}^*)$ into the Jacobian matrix can not obtain analytically. Through extensive numerical methods, we revealed that the system is unstable for any dynamical parameters. For instance, set $\mu_A=\mu_B=0.02$, $\lambda_A^\prime=\lambda_B^\prime=2.5$, $\lambda_A=0.8$ and $\lambda_B=1.5$ to compute fixed point is $(0.3344, 0.0544)$. Taking fixed point 6 into Jacobian matrix to get the two eigenvalues $\Lambda_1=-0.0048$ and $\Lambda_2=0.0040$. Thus, the fixed point is unstable.

\subsubsection{Fixed point 7}
Similar to fixed point six, we use a numerical method to analyze the stability of fixed-point seven $(\rho_A^*,\rho_B^*)=(\rho_{A_7}^*, \rho_{B_7}^*)$, and revealed that this fixed point is always unstable.

\section{Results analyses}\label{sec:Results analyses}

This section investigates the competing information spreading dynamics on simplicial complex detailedly by setting $N=1000$ nodes, the average degrees of 1-simplex and 2-simplex as $\langle k\rangle=20$ and $\langle k^\prime\rangle=6$ respectively, and the recovery rate $\mu_A=\mu_B=0.02$.

\begin{figure*}[]
\begin{center}
\epsfig{file=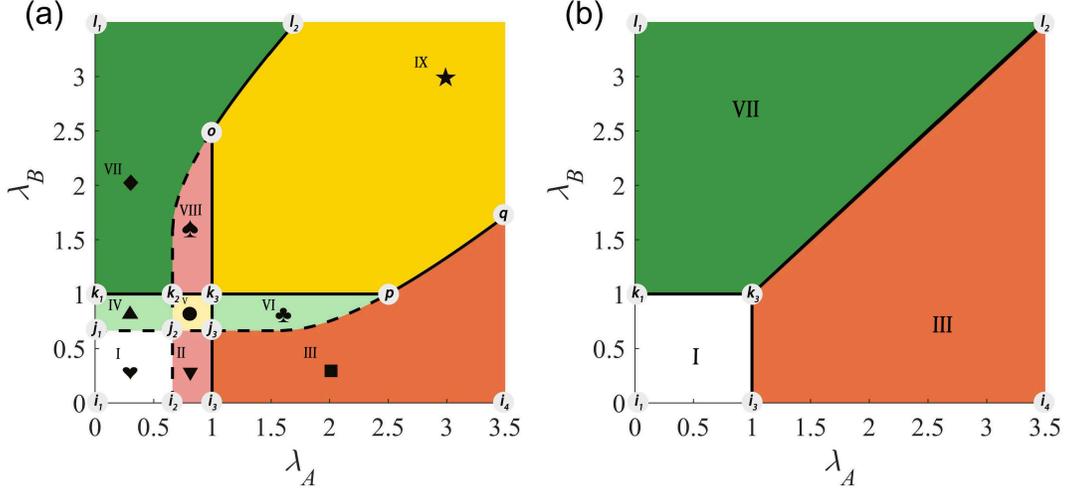,width=0.8\linewidth}
\caption{(Color online) Phase diagram of the competing epidemic spreading on simplicial complex with symmetrical interacting with $\lambda_A^{\prime}=\lambda_B^{\prime}=2.5$ (a), and $\lambda_A^{\prime}=\lambda_B^{\prime}= 0.01$ (b). In (a), the $(\lambda_A,\lambda_B)$ plane is divided into nine regions. In each region, we take a representative point to study the time evolution of the two epidemics (see Fig.~\ref{fig4}). We set the points $(0.3,0.3)$ as heart in region I, $(0.8,0.3)$ as inverted-triangle in region II, $(2,0.3)$ as square in region III, $(0.3,0.8)$ as regular-triangle in region IV, $(0.8,0.8)$ as dot in region V, $(1.6,0.8)$ as club in region VI, $(0.3,2)$ as diamond in region VII, $(0.8,1.6)$ as spade in region VIII, and $(3.0,3.0)$ as star in region IX.}
\label{fig3}
\end{center}
\end{figure*}

In Fig.~\ref{fig3}, we first show the phase diagram of the system with different values of 2-simplex infection strength, which determines the phase diagram of the system. For the case of relatively strong 2-simplex infection strength, i.e., $\lambda_{A}^\prime=\lambda_{B}^\prime=2.5$ in Fig.~\ref{fig3}(a), the $(\lambda_A,\lambda_B)$ plane is divided into nine different regions, and the phenomena are illustrated in Table~\ref{table1}. From Fig.~\ref{fig3}(a) we can extract the presence threshold and the invasion threshold:
the presence threshold $\lambda_{A}^{c_p}$ of epidemic A is $\overline{i_2j_2k_2o }$ given by Eq.~(\ref{eq:A threshold point of fp3}),
the invasion threshold $\lambda_{A}^{c_i}$ of epidemic A is $\overline{i_3j_3k_3o}$ given by Eq.~(\ref{eq:A threshold point of fp1}),
the absolute domination threshold $\lambda_A^{c_*}$ of epidemic A is $\overline{pq}$ obtained from Eq.~(\ref{eq:B threshold point of fp5}),
the presence threshold $\lambda_{B}^{c_p}$ of epidemic B is $\overline{j_1j_2j_3p}$ given by Eq.~(\ref{eq:B threshold point of fp5}), and the invasion threshold $\lambda_{B}^{c_i}$ of epidemic B line $\overline{k_1k_2k_3p}$ obtained by solving Eq.~(\ref{eq:B threshold point of fp1}), and the absolute domination threshold $\lambda_B^{c_*}$ of epidemic B is $\overline{ol_2}$ given by Eq.~(\ref{eq:A threshold point of fp3}). Since interacting parameters of epidemic A and epidemic B are symmetrical, the phase diagram is also symmetrical.

\begin{table}[!h]
\caption{ Phenomena summary of the nine regions of the phase diagram. The phenomena include dying out $(\times)$, absolute domination $(\surd)$, alternative dominance and existing hysteresis loop ($\bigcirc$). }
\label{table1}
\begin{tabular}{lccc}
\hline\hline
Regions & Epidemic A \qquad & Epidemic B \qquad & Dominant epidemic \\ \hline\hline
 I & $\times$ & $\times$ & None \\
 II & $\bigcirc$ & $\times$ & A \\
 III & $\surd$ & $\times$ & A \\
 IV & $\times$ & $\bigcirc$ & B \\
 V & $\bigcirc$ & $\bigcirc$ & A or B \\
 VI & $\bigcirc$ & $\bigcirc$ & A \\
 VII & $\times$ & $\surd$ & B \\
 VIII & $\bigcirc$ & $\bigcirc$ & B \\
 IX & $\bigcirc$ & $\bigcirc$ & A or B \\ \hline\hline
\end{tabular}
\end{table}

\begin{figure*}
 \begin{center}
 \epsfig{file=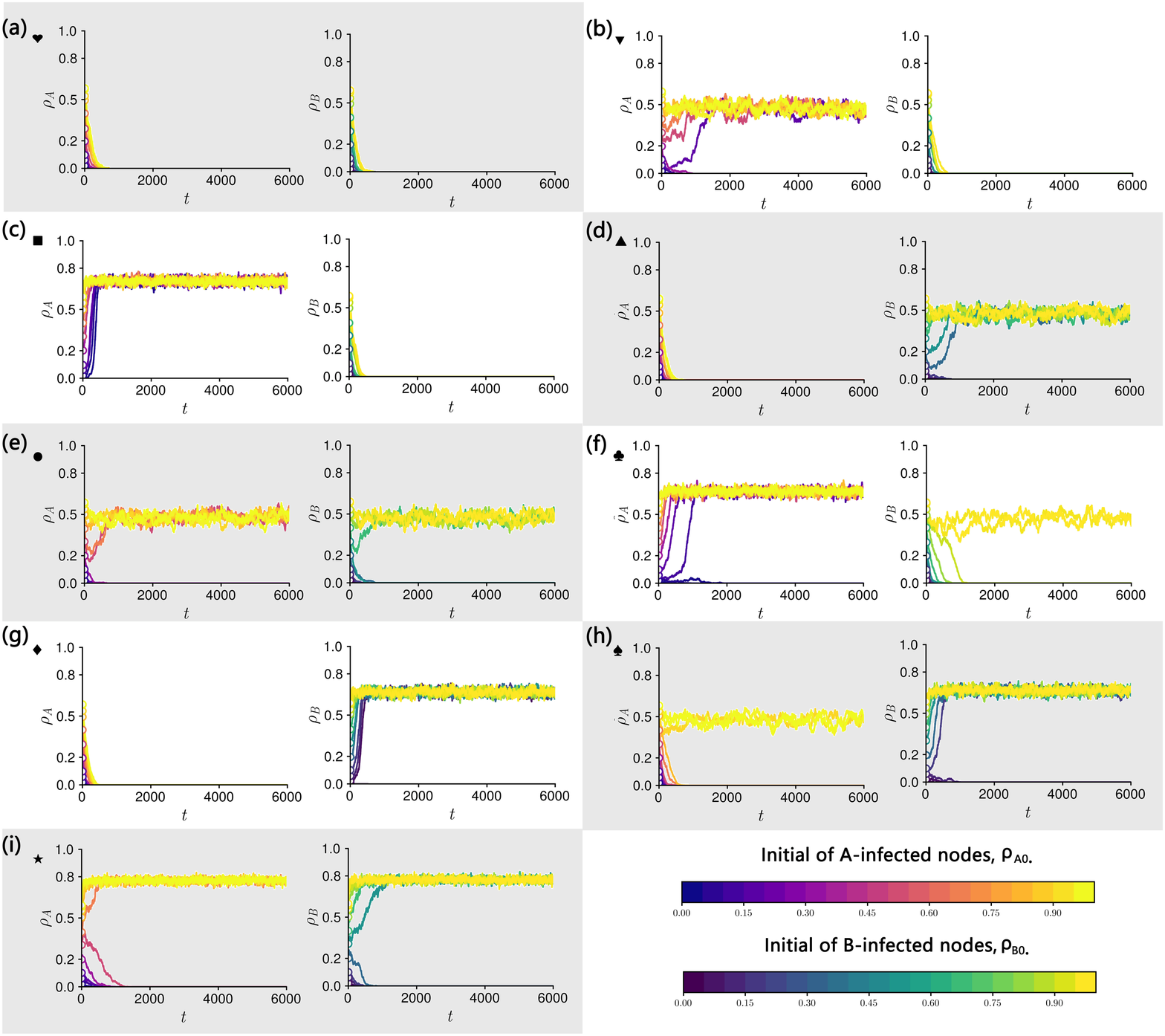,width=0.9\linewidth}
 \caption{(Color online)The time evolution of the two competing information spreading on simplicial complex when $\lambda_A^{\prime}=\lambda_B^{\prime}=2.5$. Each subfigure distributes the evolution of infected nodes fraction with time for epidemic A (left side of the subfigure) and epidemic B (right side of the subfigure) with a different fraction of initial epidemic nodes $\rho_{A0}$ and $\rho_{B0}$, in order to strengthen the influence of the initial seeds on time evolution, we set $\rho_{A0} + \rho_{B0}=0.6$. The parameters for detail of each subfigure are indicated in Fig.~\ref{fig3} (a). Subfigure (a) corresponds to the heart sign of region I in Fig.~\ref{fig3} (a). (b) corresponds to the inverted triangle of region II. (c) corresponds to the square of region III. (d) corresponds to the positive triangle of region IV. (e) corresponds to the dot of region V. (f) corresponds to the club of region VI. (g) corresponds to the diamond of region VII. (h) corresponds to the spade of region VIII. (i) corresponds to the star of region IX.}
\label{fig4}
 \end{center}
 \end{figure*}

We conduct a systematic analysis of epidemic A first.
Region I represents the epidemic-free, in which both epidemics die out, and it is determined by $\lambda_{A}^{c_p}$ and $\lambda_{B}^{c_p}$. Taking point $(0.3,0.3)$ marked by heart, for instance, the two epidemic decreases with time regardless of the values of initial seeds, and finally die out, as shown in Fig.~\ref{fig4} (a). In region II, epidemic A absolute dominates, and the hysteresis loop exists. Take point $(0.8,0.3)$ in region II marked as inverted-triangle. For instance, we study the evolution of the two epidemics in Fig.~\ref{fig4} (b). Epidemic B decreases and finally dies out. However, the survivability of epidemic A depends on the fraction of the initial node: for tiny initial seeds, epidemic A may die out in the steady-state; for large initial seeds, epidemic A will globally break out. That is to say, a hysteresis loop exists in region II.
In region III, epidemic A absolutely dominates. In Fig.~\ref{fig4} (c), we illustrate the evolutions of the two epidemics for the point $(2.0,0.3)$ marked as the square and find that epidemic A globally breaks out regardless of the initial seeds, while epidemic B dies out.
Region VIII is between the presence threshold $\lambda_{A}^{c_p}$ and the invasion threshold $\lambda_{A}^{c_i}$ of epidemic A, in which both epidemics have a hysteresis loop. Take point (0.8,1.6) in region VIII marked as a spade, for instance, the evolution of the two epidemics as shown in Fig.~\ref{fig4} (h). Epidemics A and B are alternative dominance, i.e., the two epidemics may survive, and which epidemic survive is determined by the initial seeds. Note that epidemic B is easier to survive generally than epidemic A since epidemic B can break out with a small fraction of initial seeds.

Due to the symmetry of the two epidemics, the phase transition of epidemic B is similar to epidemic A. Some regions IV, VI, and VII have similar phenomena with regions II, VIII, and III, respectively. There are two areas left in the system, regions V and IX. Similar to region VI and VIII, two epidemics hold dominance alternatively and have a hysteresis loop. Tack two points $(0.8,0.8)$ in region V and $(3,3)$ in region IX, for instance, marked as dot and star as an instance, respectively, to observe their time evolution, as shown in Figs.~\ref{fig4} (e) and (i). The survivability between the two epidemics is similar. However, in general, the survivability of the epidemics in region IX is stronger than that in region X. What is more, region V is between the presence threshold ($\lambda_{A}^{c_p}$, $\lambda_{B}^{c_p}$) and the invasion threshold ($\lambda_{A}^{c_i}$, $\lambda_{B}^{c_i}$), the epidemics are relatively fair in this region. In region IX, the epidemics are relatively fair, too. However, this region is beyond the presence threshold and the invasion threshold. Regions VI and VIII are between the two thresholds, but for regions VI, epidemic A is relatively easier to survive, and epidemic B is easier to survive in region VIII.

\begin{figure*}
\begin{center}
\epsfig{file=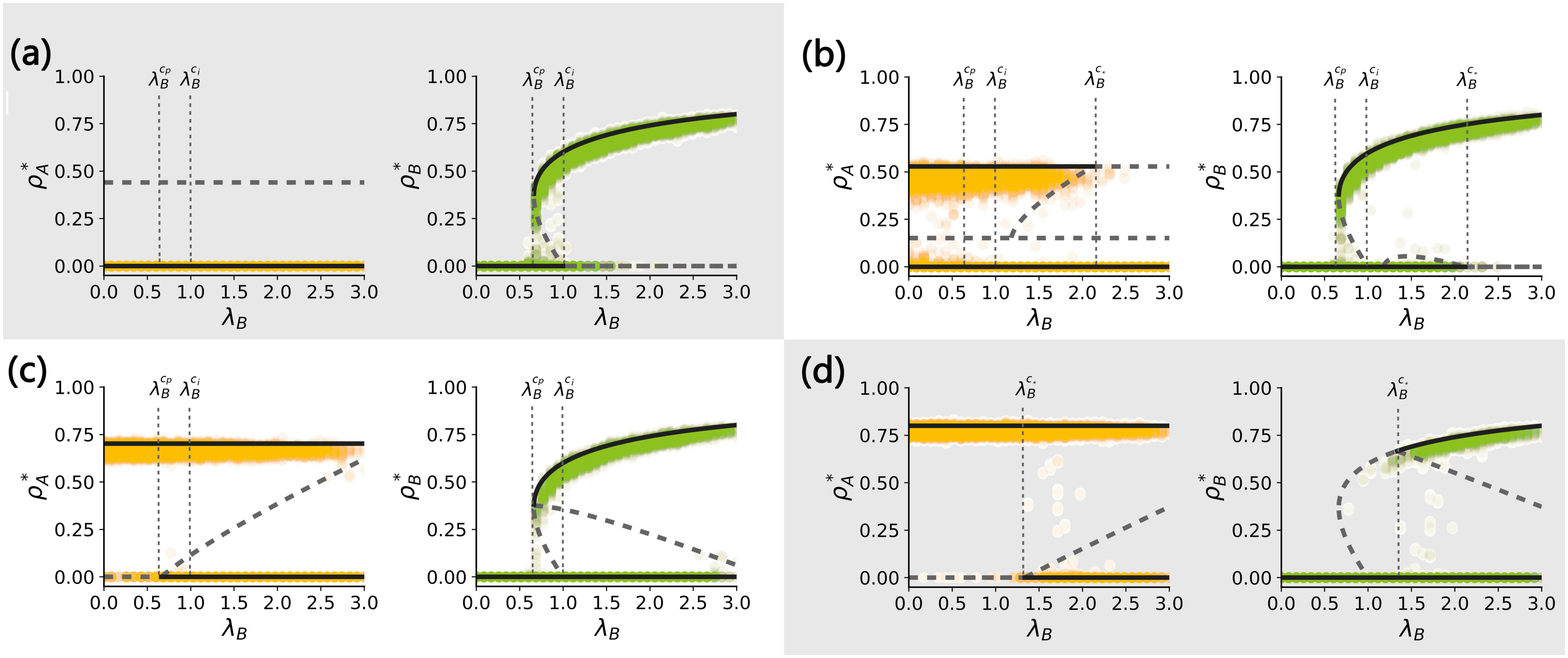,width=0.9\linewidth}
\caption{(Color online) The spreading size of the two epidemics in steady-state versus 1-simplex infection strength of epidemic B. The left side of each subfigure is epidemic A, and the right side of each subfigure is epidemic B; the vertical dashed lines indicate the presence threshold for epidemic B $\lambda_B^{c_p}$, the invasion threshold for epidemic B $\lambda_B^{c_i}$ and the threshold $\lambda_B^{c_*}$ for epidemic B absolute dominate; the solid line represents for the stable fixed point, and the dashed line represents the unstable fixed point, where the colorful dots is the simulation result of the two epidemics, orange stands for epidemic A and green stands for epidemic B. The simulation result for $\lambda_A=0.3$ in (a), $\lambda_A=0.8$ in (b), $\lambda_A=1.6$ in (c), $\lambda_A=3.0$ in (d). }
\label{fig5}
\end{center}
\end{figure*}

When the 2-simplex infection strength is extreme low, i.e., $\lambda_{A}^\prime=\lambda_{B}^\prime=0.01$ as shown in Fig.~\ref{fig3}(b),
the $(\lambda_A,\lambda_B)$ plane is divided into three regions, which is similar to the classic SIS competition model in 1-dimension networks (i.e., networks only have pairwise interactions). Region I (in white) represents the epidemic-free, region II (in dark red) represents the area of epidemic A absolute dominance, and Region VII (in dark green) represents epidemic B absolute dominance. There are no regions II, IV, V, VI, VII, and IX in this situation.

We further investigate the phase transition of epidemic B in Fig.~\ref{fig5} as the function of 1-simplex infection strength $\lambda_{B}$ for different values of $\lambda_A$. The theoretical results (i.e., lines) well agree with the simulation results (i.e., symbols).
For small values of $\lambda_A$ (e.g., $\lambda_A=0.3$) as shown in Fig.~\ref{fig5} (a), epidemic A can not break out, and epidemic B exhibits a discontinuous growth versus $\lambda_B$. There exists a hysteresis loop in the system between the presence threshold $\lambda_{B}^{c_p}$ and invasion threshold $\lambda_{B}^{c_i}=1$: for the tiny seed of epidemic B, the epidemic B can not break out (i.e., $\rho_B=0$); for the large seed of epidemic B, it globally breaks out (i.e., $\rho_B\propto o(N)$).
With the increase of $\lambda_A$, epidemic A outbreak becomes possible. When $\lambda_A=0.8$ as shown in Fig.~\ref{fig5} (b), we find the similar phenomena with Fig.~\ref{fig5} (a) when $\lambda_{B}<\lambda_{B}^{c_i}$. Note that we reveal a new phenomenon when $\lambda_{B}^{c_i}<\lambda_{B}<\lambda_{B}^{c_*}$. In this region, two epidemic alternative dominance and existing hysteresis loop. The dashed curve between $\lambda_{B}^{c_i}$ and $\lambda_{B}^{c_*}$ is the unstable fixed point six $(\rho_{A_6}^*, \rho_{B_6}^*)$ given by Eq.~(\ref{fixed_point_6}). When $\lambda_{B}>\lambda_{B}^{c_i}$ the hysteresis loop disappears, and epidemic B absolutely dominant. Different from $\lambda_A=0.8$, there is no epidemic B absolutely dominance region when $\lambda_A=1.6$ as shown in Fig.~\ref{fig5} (c), while other phenomena are still observed. For extreme strong infection strength of epidemic A (e.g., $\lambda_A=3$), epidemic A has firmly contained epidemic B, as shown in Fig.~\ref{fig5} (d), two epidemic alternative dominant and exist hysteresis loop.

\begin{figure*}
\begin{center}
\epsfig{file=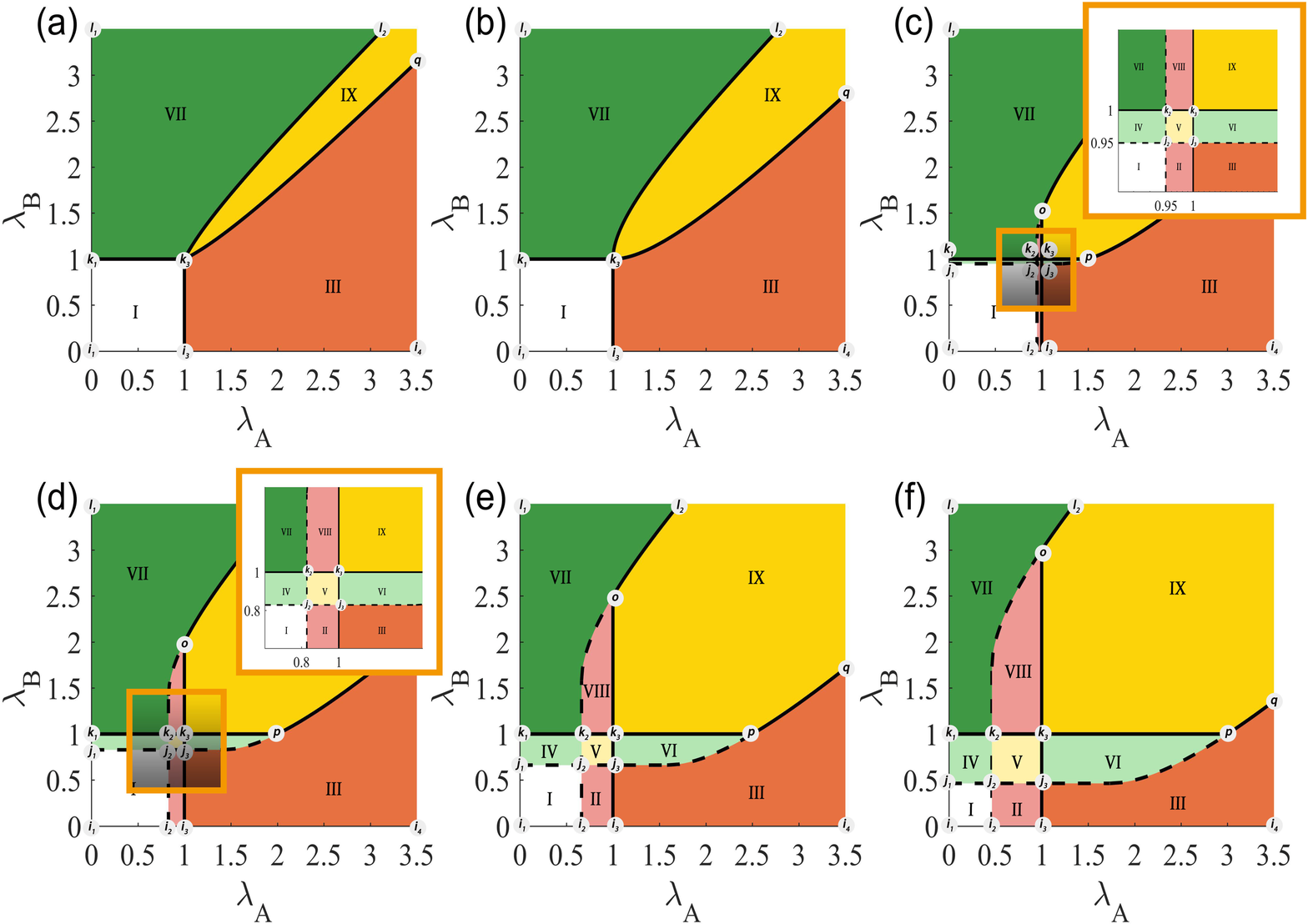,width=0.9\linewidth}
\caption{(Color online)The phase diagram under different symmetric 2-simplex infection strength with $\lambda_{A}^\prime=\lambda_{B}^\prime=0.5$ in (a), $\lambda_{A}^\prime=\lambda_{B}^\prime=1$ in (b), $\lambda_{A}^\prime=\lambda_{B}^\prime=1.5$ in (c), $\lambda_{A}^\prime=\lambda_{B}^\prime=2.0$ in (d), $\lambda_{A}^\prime=\lambda_{B}^\prime=2.5$ in (e) and $\lambda_{A}^\prime=\lambda_{B}^\prime=3$ in (f). The phenomena of regions I to IX are same as in the Table~\ref{table1}.}
\label{fig6}
\end{center}
\end{figure*}

We further explore the influence of the symmetry 2-simplex infection strength $\lambda_{A}^\prime$ and $\lambda_{B}^\prime$ on the phase diagram in Fig.~\ref{fig6}. We can see that, the area of region IX increased with the 2-simplex infection strength. However, when $\lambda_{A}^\prime=\lambda_{B}^\prime \le 1$ as shown in Figs.~\ref{fig6} (a) and (b), there are no hysteresis loop regions, i.e., regions II, IV, V, VI and VIII with the light colors are not exist in the system. Until $\lambda_{A}^\prime=\lambda_{B}^\prime > 1$ as shown in Figs.~\ref{fig6} (c)-(f) the hysteresis loop becomes more significant with the increase of $\lambda_{A}^\prime=\lambda_{B}^\prime$.

\begin{figure*}[]
\begin{center}
\epsfig{file=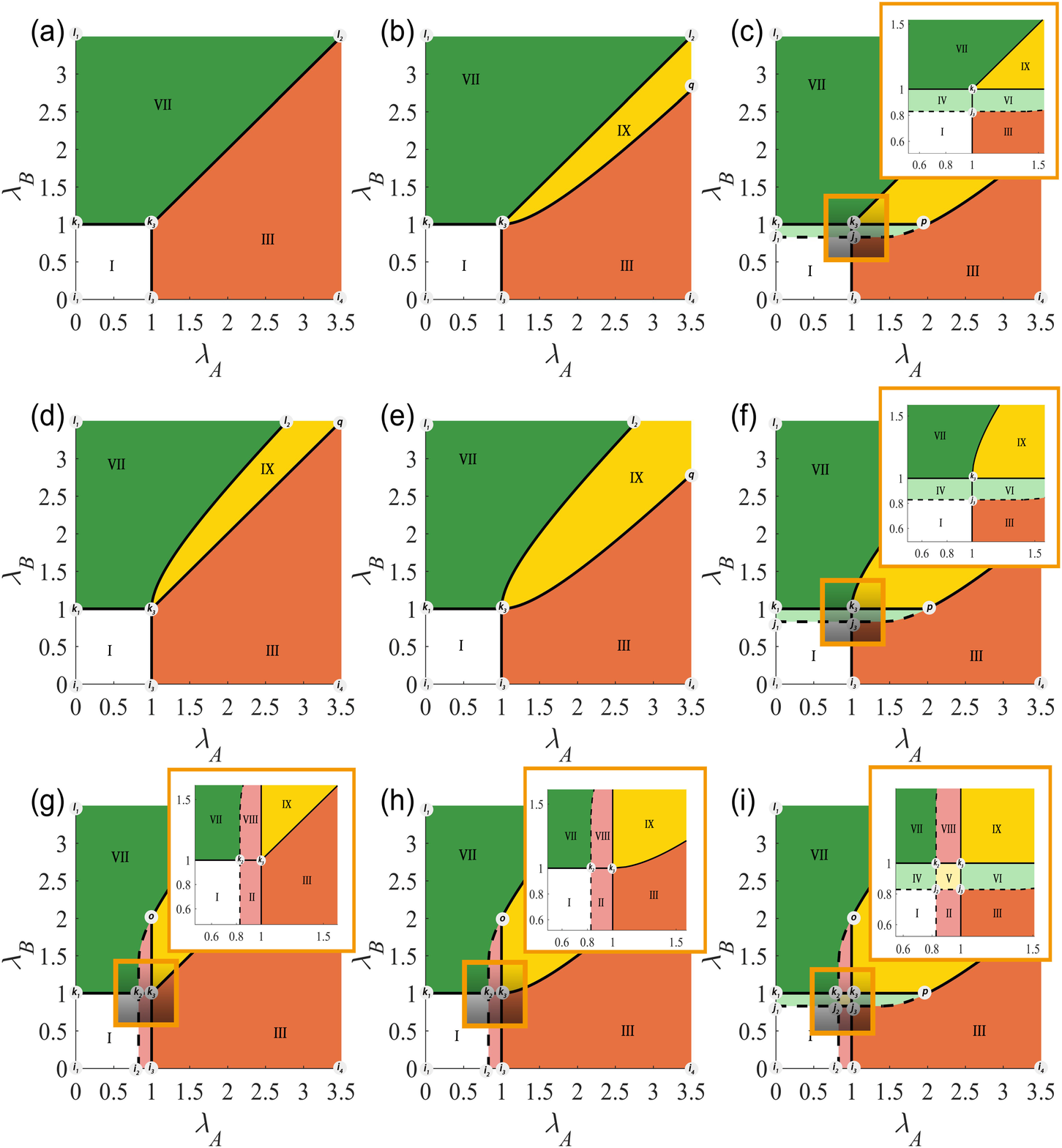,width=0.9\linewidth}
\caption{(Color online)
The phase diagram under different asymmetry 2-simplex infection strength $\lambda_{A}^\prime=0.01 $ and $\lambda_{B}^\prime =0.01 $ in (a),
$\lambda_{A}^\prime=0.01 $ and $\lambda_{B}^\prime =1.0 $ in (b),
$\lambda_{A}^\prime=0.01 $ and $\lambda_{B}^\prime =2.0 $ in (c),
$\lambda_{A}^\prime=1.0 $ and $\lambda_{B}^\prime =0.01 $ in (d),
$\lambda_{A}^\prime=1.0 $ and $\lambda_{B}^\prime =1.0 $ in (e),
$\lambda_{A}^\prime=1.0 $ and $\lambda_{B}^\prime =2.0 $ in (f),
$\lambda_{A}^\prime=2.0 $ and $\lambda_{B}^\prime =0.01 $ in (g),
$\lambda_{A}^\prime=2.0 $ and $\lambda_{B}^\prime =1.0 $ in (h) and
$\lambda_{A}^\prime=2.0 $ and $\lambda_{B}^\prime =2.0 $ in (i).
The plots in (d), (g) and (h) illustrate the situation when $\lambda_A^\prime>\lambda_B^\prime$. (b), (c) and (f) illustrate the situation when $\lambda_A^\prime<\lambda_B^\prime$. The phenomena of regions I to IX are same as in the Table~\ref{table1}.}
\label{fig7}
\end{center}
\end{figure*}

When the 2-simplex infection strength is asymmetry, the phase diagrams are shown in Fig.~\ref{fig7}, $\lambda_B^\prime$ increases from left to right, and $\lambda_A^\prime$ increases from top to bottom. We find that when $\lambda_A^\prime>1$ the hysteresis loop regions II and VIII for epidemic A exist as shown in Figs.~\ref{fig7} (g), (h) and (i). The hysteresis loop regions IV and VI for epidemic B exist when $\lambda_B^\prime>1$ as shown in Figs.~\ref{fig7} (c), (f), and (i). In addition, the area of region IX is beneficial to the epidemic with higher 2-simplex infection strength. When the infection strength of epidemic A is stronger, illustrated in Figs.~\ref{fig7} (d), (g), and (h), the distribution of region IX is beneficial to epidemic A. Similar phenomena for B epidemic are illustrated in Figs.~\ref{fig7} (b), (c) and (f).

\section{Conclusions} \label{sec:Conclusions and Discussions}
In conclusion, we have proposed a competing spreading model of two SIS-like epidemics in a simplicial complex, focusing on the influence of higher-order interactions on the critical behavior of the system. Based on the assumption that the individuals in the system are homogeneously mixed, we use the MF theory to derive the rate equations and obtain seven fixed points.
Next, by analyzing the critical conditions of the fixed points, we obtain a complex phase diagram with nine regions when the 2-simplex infection strength is significant. Region I represents the epidemic-free, in which both epidemics die out. In region II, epidemic A absolute dominates and hysteresis loop exists, the survivability of epidemic A dependents on the fraction of initial node. In region III, epidemic A absolutely dominates. In addition, both epidemics have a hysteresis loop in region VIII. They are alternative dominance, but epidemic B is easier to survive than epidemic A. Regions IV, VI, and VII have similar phenomena with regions II, VIII, and III, respectively, due to the symmetry of the two epidemics. Regions V and IX are alternative dominant regions, where both epidemics have a hysteresis loop.
However, when the 2-simplex infection strength is extremely weak that it can be ignored, the phase diagram of the system is consistent with the one in the simple graph. Next, combining the theory with many numerical simulations, we explained the time evolution and steady-state outbreak size of the two epidemics in each region. Moreover, the theoretical outbreak size matches the simulation well.
Furthermore, we explored the phase diagram when the 2-simplex infection strength is symmetrical and asymmetrical. The results show that the 2-simplex infection strength has a significant impact on the system phase diagram. The existence of regions II V and VIII (or regions IV V and VII) are related to whether the 2-simplex infection strength of epidemic A (or B) is greater than one, and when these regions exist, the size of these regions are positively correlated with the 2-simplex infection strength. We can see that the existence and size of region V are related to the 2-simplex infection strength of both epidemics. Moreover, region IX always exists in the system, the area of region IX is beneficial to the epidemic with higher 2-simplex infection strength.
We have worked the simplest model of the competing dynamics on a higher-order system. The method applied here is based on the ideal MF theory and has some differences from reality. Therefore, some more accurate theories need to be further studied, such as heterogeneous MF, Microscopic Markov chain approach. Nonetheless, our research provides a specific theoretical basis for competition models in higher-order interactions and helps explain complex competition phenomena in the real world.

\acknowledgments

\noindent
This work was partially supported by the National Natural
Science Foundation of China under Grants No.~61903266
and Sichuan Science and Technology Program (No. 2020YJ0048).


\end{document}